\documentclass[]{osa-article}

\journal{oe}
\usepackage{color}
\definecolor{mygreen}{rgb}{0,0.5,0}
\definecolor{mygrey}{rgb}{0.5,0.5,0.5}
\definecolor{myred}{rgb}{0.75,0,0}
\definecolor{myblue}{rgb}{0,0,0.75}
\definecolor{mymagenta}{cmyk}{0,1,0,0.12}
\definecolor{mycyan}{cmyk}{1,0,0,0.12}
\definecolor{myorange}{rgb}{1.,0.5,0}
\definecolor{myviolet}{rgb}{0.6,0.15,0.6}
\definecolor{mybrown}{cmyk}{0,0.50,1,0.41}

\usepackage{ulem}

\definecolor{darkgreen}{RGB}{0,170,50}

\newcommand{\subdk}{_{\mathrm{dk}}}
\newcommand{\subrt}{_{\mathrm{rt}}}
\newcommand{\subcoll}{_{\mathrm{coll}}}
\newcommand{\subwd}{_{\mathrm{wd}}}
\newcommand{\subse}{_{\mathrm{se}}}
\newcommand{\subsd}{_{\mathrm{sd}}}
\newcommand{\subbg}{_{\mathrm{bg}}}
\newcommand{\subexp}{_{\mathrm{exp}}}

\usepackage{hyperref}

\usepackage{siunitx}
\DeclareSIUnit\torr{Torr}
\DeclareSIUnit\amagat{amg}

\articletype{Research Article}

\begin{document}

\title{Laser-written vapor cells for chip-scale atomic sensing and spectroscopy}

\author{Vito G. Lucivero,\authormark{1*} Andrea Zanoni,\authormark{2,3} Giacomo Corrielli,\authormark{2} Roberto Osellame,\authormark{2} and Morgan W. Mitchell\authormark{1,4}}

\address{\authormark{1}ICFO - Institut de Ci\`encies Fot\`oniques, The Barcelona Institute of Science and Technology, 08860 Castelldefels (Barcelona), Spain\\
\authormark{2}Istituto di Fotonica e Nanotecnologie (IFN) — Consiglio Nazionale delle Ricerche (CNR), Piazza Leonardo da Vinci 32, 20133 Milano, Italy\\
\authormark{3}Dipartimento di Fisica — Politecnico di Milano, Piazza Leonardo da Vinci 32, 20133 Milano, Italy\\
\authormark{4}ICREA - Instituci\'{o} Catalana de Recerca i Estudis Avan{\c{c}}ats, 08010 Barcelona, Spain\\
}

\email{\authormark{*}vito-giovanni.lucivero@icfo.eu} 



\begin{abstract}

We report the fabrication of alkali-metal vapor cells using  femtosecond laser machining. This laser-written vapor-cell (LWVC) technology allows arbitrarily-shaped 3D interior volumes and has potential for integration with photonic structures and optical components. We use non-evaporable getters both to dispense rubidium and to absorb buffer gas. This enables us to produce cells with sub-atmospheric buffer gas pressures without vacuum apparatus. We demonstrate sub-Doppler saturated absorption spectroscopy and single beam optical magnetometry with a single LWVC. The LWVC technology may find application in miniaturized atomic quantum sensors and frequency references.
\end{abstract} 
\section{Introduction}
Since its first demonstration in 1994 \cite{du1994}, femtosecond-laser-writing (FLW) has emerged as a very versatile, direct and maskless technique for inscribing bulk optical waveguides as well as buried microchannels in transparent materials \cite{gatt08} and it has been applied to numerous kind of substrates including glasses, ceramics, polymers and crystals \cite{osel12, chen14}. Thanks to its unique 3D structuring capabilities, FLW has enabled many innovative geometries for the fabrication of integrated photonic devices like directional couplers and Y-splitters\cite{della08}, complex multi-paths interferometers \cite{crespi13}, 3D waveguide lattices \cite{corr13,rech13}, polarization rotators \cite{Corrielli2014,heil14}, and Bragg reflectors \cite{zhan07}. This versatility, in addition to low-cost and rapid prototyping, has allowed the fabrication of FLW-based devices for waveguide-assisted applications in several fields like astrophotonics \cite{Thomson2009}, telecommunications \cite{Eaton2009}, high-order harmonic generation in gases \cite{Ciriolo2020}, optical \cite{Zhang2008, mart17} and NV center-based sensing \cite{Hoese2021}. Most notably, FLW-based optofluidic and microfluidic lab-on-chip techniques, in which three-dimensional microchannels and optical waveguides are monolitically integrated within the same substrate, \cite{Sugioka2005,osel11} are now a mature technology used world-wide for chemical analysis \cite{yin21, wu21}, biosensors \cite{Crespi2010} and single-cell processing tools \cite{sala20, paie14, brag12}. More recently, FLW has played an important role also in the fabrication of photonic integrated circuits (PICs) for use in quantum technologies including photonic quantum computing and simulation, quantum communication and quantum sensing \cite{Corr21,Meany2015}. Low waveguide losses and birefringence, the excellent connectivity of laser-written PICs with standard optical fibers and the possibility of fabricating integrated quantum memories \cite{Seri2018} make FLW attractive for these applications.\par
Meanwhile, in recent years there has been growing interest in the development of chip-scale and miniaturized atomic sensors \cite{Kitching2018}. Fabrication techniques based on micromachined structures in silicon, which usually employ anodic bonding between Si and Borofloat glass substrates to make MEMS vapor cells \cite{Knapkiewicz2019}, have shown great scalability and stability properties. Alternatively, photolithography of a photoresist, in combination with etching and glassblowing techniques, has been used to fabricate atomic microchannels \cite{Baluktsian2010} and, more recently, nanostructured alkali-metal vapor cells \cite{Peyrot2019,Cutler2020}. However, due to the planar bonding or lithographic strategy, all these cells do not show 3D versatility and have limited optical access. Many atomic sensors like those based on multipass cells \cite{Limes2020,Lucivero2021,Lucivero2022}, cell arrays \cite{IJsselsteijn2012}, dual-beam \cite{Kominis2003}, triple-beam \cite{Boudot2020,Lee2021} or cavity-enhanced \cite{Crepaz2015,Mazzinghi2021} geometries, do require greater flexibility and optical access. \\ Various strategies have been also developed for the integration of vapor cells with photonic waveguides. The first one is based on the interaction between hot vapors and the evanescent field of either photonic waveguides \cite{Ritter2015,Tombez2017} or tapered optical fibers \cite{Hendrickson2010,Takiguchi2011} down to the nanoscale \cite{Stern2013}. A second one is based on atomic filling of antiresonant ARROW \cite{Yang2007} or hollow core photonic fibers \cite{Knabe2009,Slepkov2010,Lurie2012,Triches2015}. While the tight confinement on the micro- or nano-scale has been used to demonstrate strong light-atom interaction for enhanced nonlinear effects \cite{Spillane2008,Stern2017} and new optical phenomena \cite{Keaveney2012,Petersen2014,Lodahl2017,Peyrot2018}, the broadening due to rapid transit time or walls collisions limits its utility for high-sensitivity atomic sensors and precise atomic spectroscopy \cite{Kitching2018}. Despite of the mitigation of this effect in hollow core fibers, which can reach mode diameters as large as 85 $\mu$m \cite{Perrella2012}, this strategy also suffers from frequency shifts due to the multi-mode nature of the guiding fibers \cite{Light2015} and lack of geometry versatility. In general, an atomic interaction size from 100 $\mu$m to 1 mm, i.e. typically set by the transverse beam width, is desirable for Doppler-free applications \cite{Kitching2018}, like atomic frequency references \cite{Knappe2007} and atomic clocks \cite{Newman2019,Micalizio2021}, as well as for high-sensitivity spin-based atomic sensors like optical magnetometers \cite{Budker2007,Lucivero2014} and atomic gyroscopes \cite{Kornack2005}. The combination of the latter goal with integrated photonics led to state-of-the-art photonic-atomic chips, which make use of hybrid schemes where a separate mm or $\mu$m-sized microfabricated vapor cell is deposited on the top of the photonic chip \cite{Hummon2018,Sebbag2021}. Then, complex extreme mode-converting apodized grating structure or a nanophotonic spin selector are used to couple light between atomic microcell and photonic waveguides, respectively \cite{Hummon2018,Sebbag2021}, with relatively low efficiency.\par
In this work we introduce the fabrication of laser-written vapor cells (LWVCs) on fused silica to extend the range of FLW applications to atomic spectroscopy and spin-based sensing. After describing the LWVCs fabrication and characterization, we demonstrate proof-of-principle applications as saturated absorption spectroscopy and optically-pumped magnetometry with a rubidium vapor. The introduced technique provides: 1) maskless fabrication of alkali-metal vapor cells with 3D versatility of buried atomic channels and reservoir; 2) suitable interaction volume for sub-Doppler spectroscopy and high-sensitivity atomic quantum sensors;  3) high potential for scalability and integration with waveguide-based PICs. Furthermore, we show that the use of non-evaporable getters (NEG) technology can release a Rb vapor and, at the same time, strongly reduce buffer gas pressure with no need for a vacuum apparatus.  We note that some vapor cell applications are sensitive to alkali diffusion into the host material \cite{MaPRA2009,Karlen2017}, or to the proximity of residual alkali metal droplets to the optically probed region \cite{GriffithOE2010}. The ability of FLW to work with hard materials such as fused silica, and to create arbitrary cavity geometries, will be advantageous in such scenarios.


\section{Vapor cell fabrication}

\subsection{Laser writing of channels and reservoir}

\begin{figure}[t!]
\centering
\includegraphics[width=12cm]{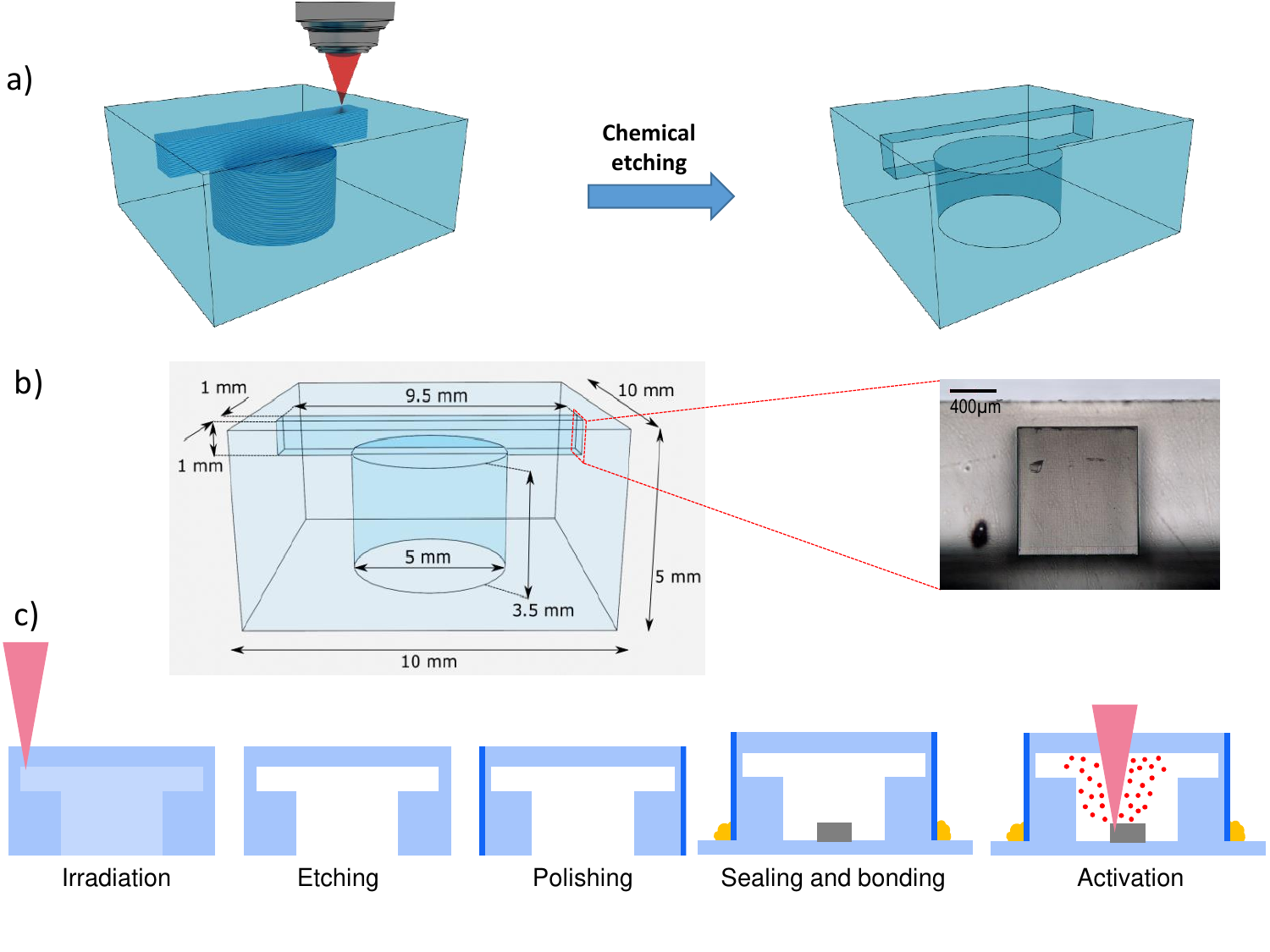}
\caption{(a) Schematic of the FLICE process adopted for the fabrication of the LWVC. Ultrafast laser writing is used to define the geometry of the cell and the rubidium reservoir inside the bulk of a fused silica substrate. Then, the substrate is exposed to chemical etching and the irradiated material is removed. (b) Layout of the fabricated LWVC. In the inset, a microscope picture of the lateral facet of the sensing channel is shown. (c) Complete workflow of the LWVC fabrication process with 2D cross-section at each step.}
\label{fig:Fabrication}
\end{figure}

For the fabrication of the LWVC we employed the so-called FLICE (Femtosecond Laser Irradiation followed by Chemical Etching) technique. This is a two-step microfabrication process illustrated in Fig.~\ref{fig:Fabrication}(a). At first, an ultrafast laser beam is tightly focused inside the volume of a fused silica glass substrate. For a suitable choice of the irradiation parameters, this step leads to the local formation of self-oriented nanogratings, which, in turn, produces an increased sensitivity of the irradiated volume to chemical etching. Thanks to the non-linear nature of the laser absorption process, the material modification is highly localized within the laser focal spot, thus providing extreme flexibility in the inscription of an arbitrary 3D pattern buried inside the substrate. In a second step, the substrate is immersed in a proper etchant solution, and the irradiated material is selectively removed, leading to the formation of an empty structure with the desired geometry.

To fabricate the first LWVC, used in the experiments described in section \ref{section:experiments}, we performed the irradiation step using a commercial femtosecond laser source (HighQ Laser - femtoREGEN), that emits a train of ultrafast light pulses at 520~nm (300~fs duration, 410~nJ/pulse) at the repetition rate of 1~MHz. A $50\times$ microscope objective with 0.6 numerical aperture is used to focus the laser output inside a pristine fused silica block of dimensions $\SI{10}{\milli\meter}\times\SI{10}{\milli\meter}\times\SI{5}{\milli\meter}$. The irradiated geometry, illustrated in Fig.~\ref{fig:Fabrication}(b), consists of a cylinder oriented vertically with a diameter of 5~mm and a height of 3.5~mm, which forms the reservoir for the rubidium dispenser, and a rectangular parallelepiped inscribed on top to it, with a $\SI{1}{\milli\meter}\times\SI{1}{\milli\meter}$ cross section and a length of 9.5~mm, which forms the atomic sensing chamber. The top facet of the reservoir and the bottom facet of the chamber are in direct contact for facilitating the material removal during the etching step. The chamber is completely buried within the glass block, while the reservoir reaches its bottom facet, allowing the penetration of the acid and filling of the LWVC. The laser definition of these shapes is limited to their surfaces, with additional section planes written both in the reservoir and in the chamber. This is enough to cause the complete detachment of the inner material during the etching step. The overall structure has been irradiated bottom-to-top, with a substrate scan speed of~1 mm/s and a vertical line separation of 3~$\mu$m. During the whole process, the laser polarization was kept oriented perpendicular to the longitudinal axis of the parellelepiped. In this way, the nanogratings resulted aligned parallel to this direction, thus favouring the acid penetration and increasing the etching speed of the sensing chamber. Importantly, when more complex structures are fabricated by FLICE, a proper polarization direction must be chosen during the irradiation of each sub-component, in order to temporally synchronize the etching of the whole geometry.

The chemical etching of the irradiated structure was performed by dipping the fused silica block in a \SI{20}{\percent} hydrofluoric acid aqueous solution, at the temperature of~35~°C. The whole etching process lasted $\approx\SI{18}{\hour}$. The final distance between the sensing chamber and the top surface of the glass substrate is $\approx \SI{300}{\micro\meter}$, as shown in the inset of Fig.~\ref{fig:Fabrication}(b). This distance could be potentially reduced to few tens of microns, thus combining the ability of writing buried sensing channels with a sub-mm stand-off distance from a potential sample on the top surface. Finally, once the LWVC was fully etched, the lateral facets of the glass block, parallel to the sensing channel cross section, have been manually polished in order to reduce the light scattering losses during the sensing experiments. The inner surfaces of the fabricated structure corresponding to the input and output facets of the sensing chamber present a residual roughness in the range 50 nm - 100 nm RMS \cite{Turco17,Dogan20}, which introduces additional scattering losses measured to be around \SI{8}{\percent} per rough facet. Note that surface roughness can be greatly reduced (down to \SI{20}{\nano\meter} RMS) by performing an additional thermal annealing process on the sample after the etching step \cite{Sala2021}.

\begin{figure}[t!]
\centering
\includegraphics[width=12cm]{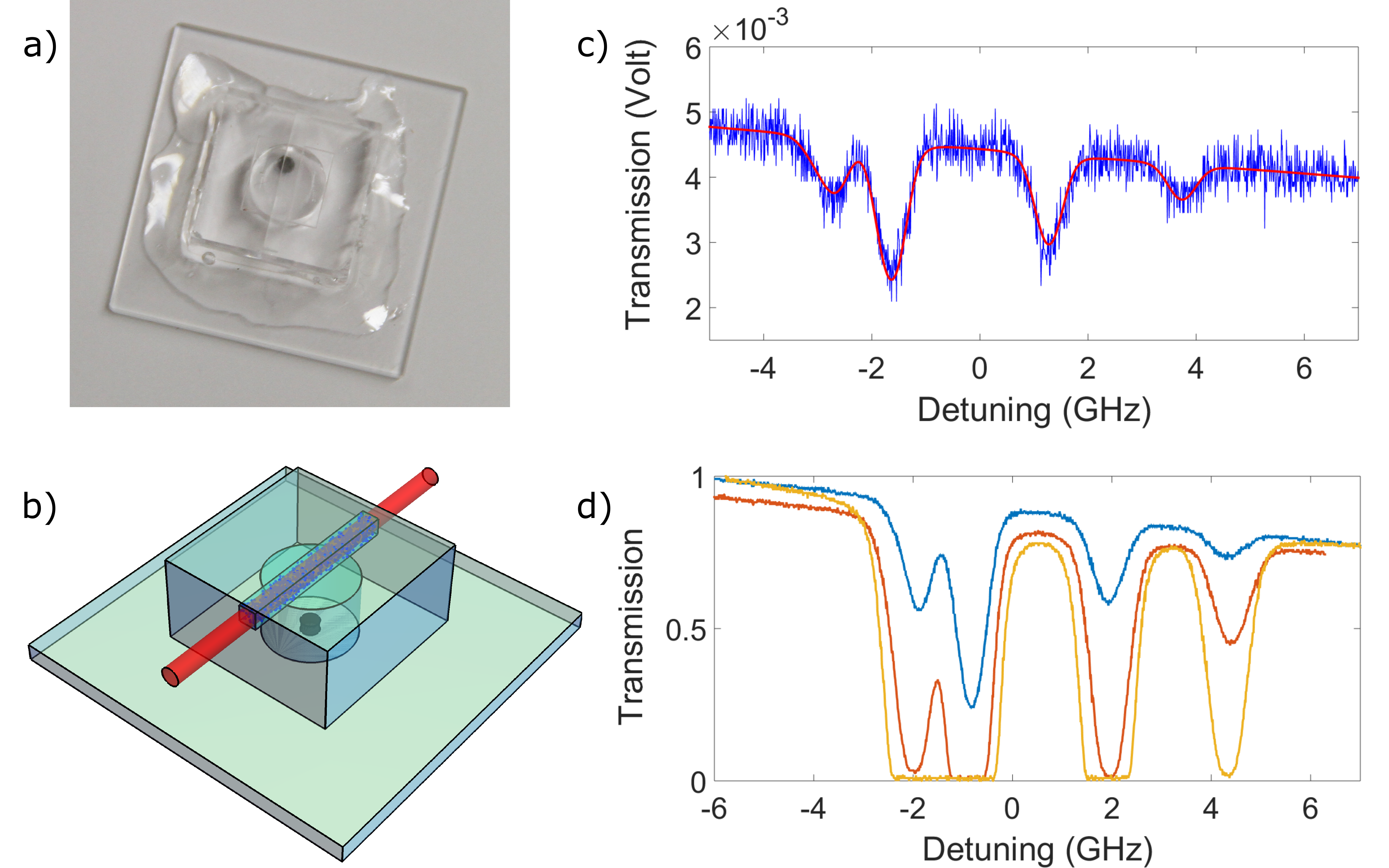}
\caption{\textbf{a)} Laser-written vapor-cell (LWVC) after filling with a solid state Rb dispenser pill and bonding with UV curing glue to a bottom plate. \textbf{b)} 3D sketch of the LWVC with a probe beam propagating through the sensing channel with atoms (blue dots) in vapor phase. \textbf{c)} Absorption spectrum (blue) and Voigt fit (red) with the cell at $T=\SI{45}{\celsius}$ in the weak-probe regime.  \textbf{d)} normalized absorption spectra at temperatures of $T=\SI{60}{\celsius}$ (blue), $T=\SI{70}{\celsius}$ (red) and $T=\SI{80}{\celsius}$ (orange). Zero detuning corresponds to the $^{87}$Rb D2 line center.} 
\label{fig:LWVCFilled&Spectra}
\end{figure}

\subsection{Device filling and bonding}
When the FLICE process is completed, we proceed with the device filling and bonding. The complete workflow of the LWVC fabrication process is depicted in Fig.~\ref{fig:Fabrication}(c). At first, we place a pill of non-evaporable getter (NEG) material (SAES
Getters RB/AMAX/PILL/1-0.6)  in the reservoir cavity of the fused silica structure. As shown in Fig. \ref{fig:LWVCFilled&Spectra}(a), we use UV-curing epoxy to bond a \SI{1}{\milli\meter}-thick fused silica plate to seal the LWVC. This sealing is performed in a chamber filled with atmospheric pressure N\textsubscript{2} containing less than \SI{6}{ppm} of O\textsubscript{2}, to avoid trapping O\textsubscript{2} in the cell.  As we describe in section 3, UV-curing epoxy is convenient and suitable for atomic sensing applications. Furthermore, the FLICE-based fabrication of vapor cells is also compatible with other sealing methods like optical contact bonding \cite{Peyrot2019,Cutler2020}, anodic bonding \cite{DuralPatent} and laser microwelding \cite{Wlodarczyk2019}, for more demanding requirements of utra-high vacuum and temperatures above \SI{200}{\celsius}.
The getter material, when heated, both releases Rb and consumes N\textsubscript{2}, working as a passive getter pump with no need for vacuum apparatus. Specific buffer gas pressures can in principle be achieved post-sealing, either by controlling the duration of the getter heating, so as to limit the consumption of buffer gas, or by filling the cell with a controlled fraction of ungetterable gases, e.g. Ar, followed by a strong heating of the getter material.  For cold atoms applications, a pressure lower than \SI{2e-7}{\torr} has been recently reported using NEG dispensers \cite{Little2021}. We note also that the fused silica has a broad optical transparency window, which may be convenient for spectroscopic identification of the precise composition of the gas content of the cell.

We activate the getter with a \SI{6}{\watt} beam of cw  \SI{1064}{\nano\meter} light, focused to a \SI{200}{\micro\meter} waist at the dispenser. Illumination for \SI{20}{\second} was sufficient to release abundant Rb for spectroscopic purposes and also to consume most of the N\textsubscript{2} in the chamber \cite{mauricePhDThesis}. 

We characterize the LWVC's optical quality and gas fill after activation by single-pass absorption spectroscopy. The LWVC is enclosed in a ceramic, resistively-heated oven, and light from a distributed Bragg reflector (DBR) laser is passed through the LWVC's long chamber, as illustrated in Fig. \ref{fig:LWVCFilled&Spectra}(b). At room temperature, and thus low vapor density, a transmission of \SI{50}{\percent} is observed. We attribute this transmission to scattering losses due to non-ideal alignment of the light beam with the chamber axis and to the residual roughness of its inner facets. With a low probe intensity of about \SI{1}{\micro\watt\per\milli\meter^2}, chosen to avoid optical pumping and line-broadening effects, transmission spectra show clearly all features of the Rb D2 line, see Fig. \ref{fig:LWVCFilled&Spectra}(c).  We fit the absorption spectrum at T$=\SI{45}{\celsius}$ with a Voigt profile and infer a pressure-induced broadening of about \SI{70}{\mega\hertz}, corresponding to a N$_2$ residual pressure of \SI{3.5}{\torr} \cite{Romalis1997}. This  condition, in which collisional broadening is between the $6$ MHz natural linewidth and the $500$ MHz Doppler broadening, is suitable to demonstrate sub-Doppler saturated spectroscopy.  This pressure also significantly slows the atomic motion, thus extending the spin coherence time, of importance for spin-based atomic sensing. 

\begin{figure*}[t!]
\centering
	\includegraphics[width=12cm]{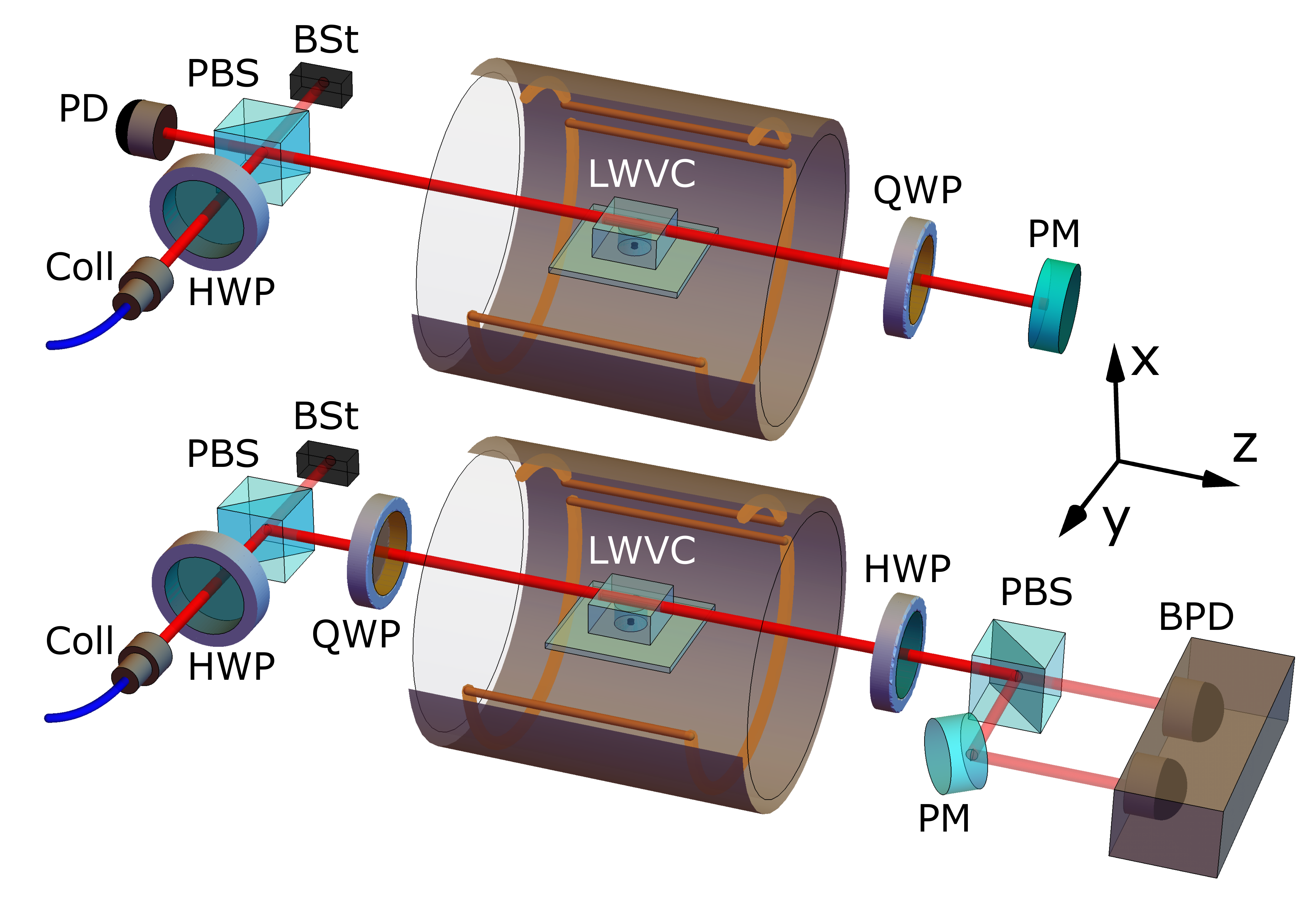}
\caption{\textbf{Experimental setup.} The LWVC stands within a layer of $\mu$-metal shielding and a system of concentric coils. The laser beam is coupled to a fiber collimator, the power reaching the LWVC is adjusted with a half-wave-plate (HWP) and a polarizing-beam-splitter (PBS), while the residual power is absorbed by a beam stop (BSt). \textbf{(Top) Saturated absorption spectroscopy (SAS).} The SAS setup is a double-pass configuration including a quarter-wave-plate (QWP) after atomic interaction, a fully reflecting planar mirror (PM) and a photo-detector (PD) after double-pass through the LWVC. \textbf{(Bottom) Optical magnetometry.} The optical magnetometer setup is a single-pass configuration including a QWP before atomic interaction and a polarimeter, which consists of HWP, PBS, PM and an amplified differential photo-detector (BPD).}
\label{fig:ExpSetup}
\end{figure*}

\section{Atomic sensing with LWVCs}

We now describe two representative sensing tasks: saturated absorption spectroscopy (SAS) and single-beam optically-pumped magnetometry (OPM), implented with a LWVC. Achieving competitive performance in either of these application \cite{Hummon2018,Shah2009} requires considerable technical effort and is beyond the scope of this manuscript. Nonetheless, their implementation  shows the  compatibility of the LWVC methods with these applications and with related techniques such as spin noise spectroscopy \cite{Swa2018,LuciveroSNS2016}. In the OPM case, we also compare against models of diffusion- and collision-limited spin coherence times and find good agreement.  

\subsection{Experimental setups}
\label{section:experiments}
We use the experimental apparatuses shown in Fig. \ref{fig:ExpSetup} to perform SAS, suitable for laser frequency stabilization, and single-beam optical magnetometry, as a paradigmatic
application of atomic quantum sensing. The LWVC is enclosed in a magneto-optical setup which consists of a single layer of magnetic shielding and coils to generate dc and gradient fields. The LWVC is housed in a ceramic oven, intermittently heated with resistive heaters and temperature stabilized to \SI{0.1}{\celsius} using a thermocouple sensor. A laser beam is fiber coupled and the power is adjusted through a half-waveplate and a polarizing beam splitter, which also ensures linear polarization of the reflected beam reaching the LWVC for atomic interaction. 

\subsection{Saturated absorption spectroscopy}
\begin{figure}[h!]
\centering
	\includegraphics[width=12cm]{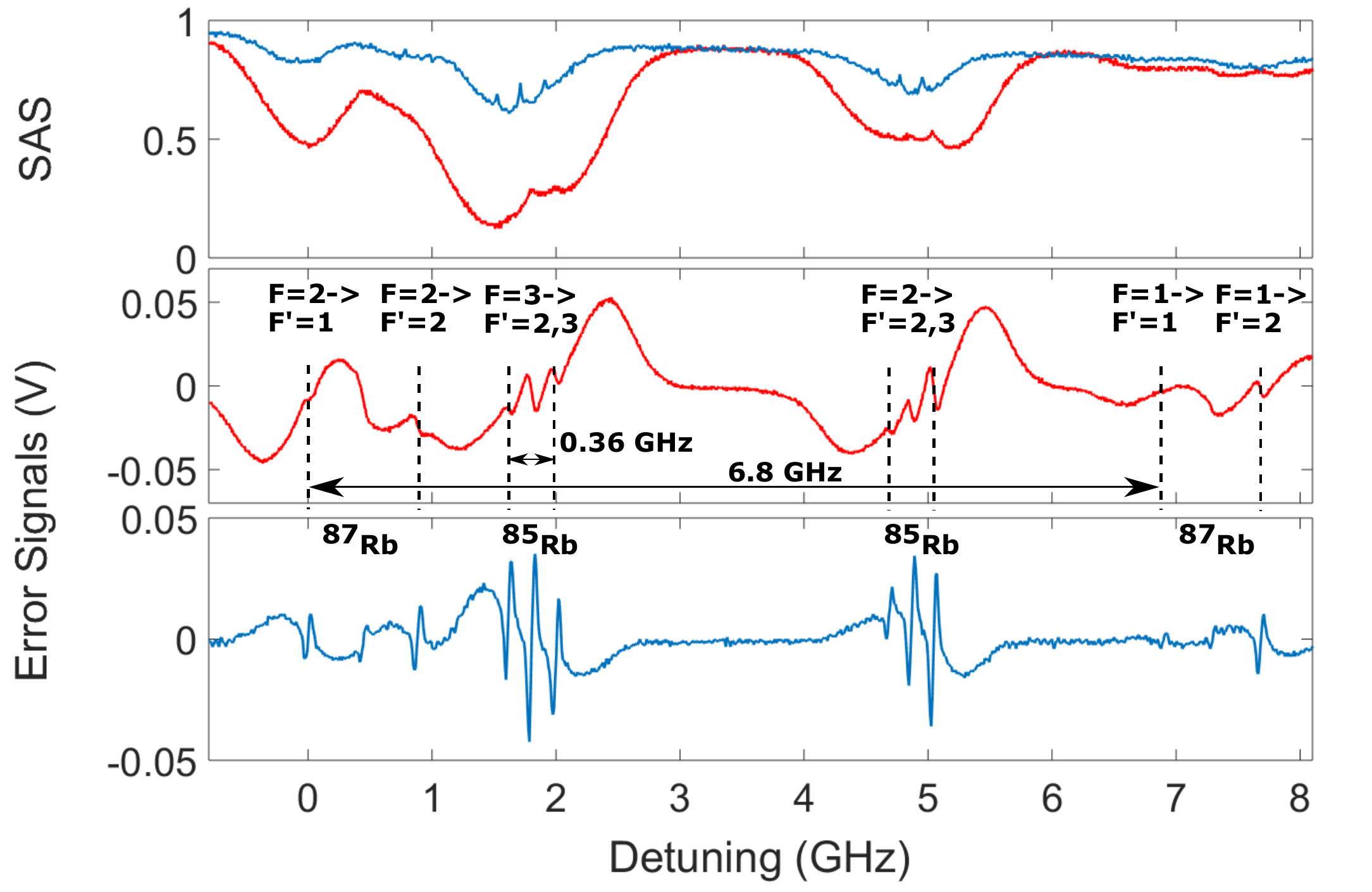}
\caption{\textbf{SAS Signals.} \textbf{Top:} D1 line Rubidium saturated absorption spectra for the LWVC (red) and the vapor reference cell (blue). \textbf{Middle:} error signal for RF spectroscopy with the LWVC at 70 $^{\circ}$C. \textbf{Bottom:} error signal with the table-top vapor reference cell at room temperature. The zero detuning corresponds to the F=2->F`=1 resonance of $^{87}$Rb.}
\label{fig:SAS}
\end{figure}

As shown in Fig. \ref{fig:ExpSetup} (top), we perform SAS by sending \SI{1}{\milli\watt} of a linearly-polarized \SI{795}{\nano\meter} laser beam (Toptica DL100), tunable around the Rb D1 transition, through the laser-written channel, retro-reflecting with a planar mirror, and detecting the transmitted light with a \SI{150}{\mega\hertz} amplified photodetector.  A quarter-wave plate before the mirror flips the polarization between the two passes.  The laser current is modulated at \SI{20}{\mega\hertz} to produce frequency modulation (FM) of the probe, and the detected photocurrent is demodulated to recover an error signal proportional to the derivative with respect to frequency of the transmission.  The same technique is simultaneously applied, with the same laser power, to obtain the error signal from a commercial Rb cell with no buffer gas and a \SI{71.8}{\milli\meter} internal length (not shown in Fig. \ref{fig:ExpSetup}).  Fig. \ref{fig:SAS} shows  SAS spectra for both the LWVC stabilized at 70 $^{\circ}$C and the conventional reference cell at room temperature. These resolve all sub-Doppler and crossover resonances for both $^{85}$Rb and $^{87}$Rb isotopes. 
We note that the SAS features (the narrow absorption dips) are, due to pressure broadening, about ten times broader than the natural linewidth. A lower buffer gas pressure, either by sealing in vacuum or by further gettering by the dispenser material, can be expected to reduce this width and give a corresponding boost in SNR.

\subsection{Optical magnetometer}
\begin{figure}[h!]
	\includegraphics[width=\textwidth]{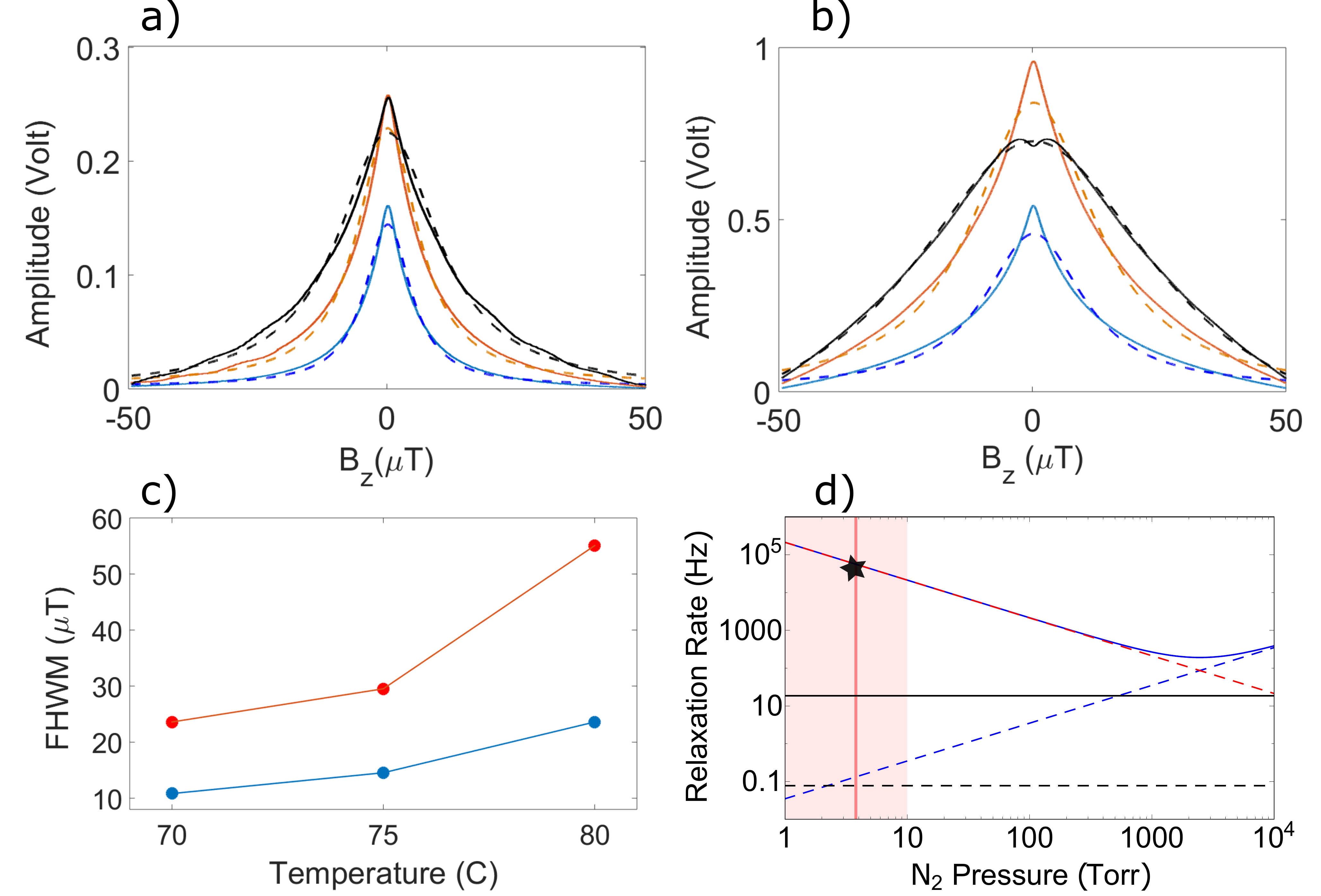}
\caption{
\textbf{a-b) Zero-field resonance magnetometry signals.} a) and b) show rotation signals versus transverse field for D2 and D1 lines, respectively, acquired with \SI{3.5}{\milli\watt} of probe power. Solid lines show experimental data, dashed lines show prediction of Eq. (\ref{Eq:signal}). Blue, orange and black lines show rubidium number density $n = (0.74, 1.0, 1.5) \times 10^{12}$ atoms/cm$^3$, respectively, corresponding to cell temperatures of $T=(70, 75, 80)^{\circ}\mathrm{C}$, respectively. \textbf{c) Resonance linewidth.} Experimental full-width-half-maximum (FWHM) linewidth $2\times\Delta B_x^{\rm{exp}}$ versus temperature for D1 (red) and D2 (blue) ZFR signals as in a) and b). \textbf{d) Relaxation rates}. Calculated ``in the dark'' relaxation rate $\Gamma\subdk^{\rm{ th}}/2\pi$ (blue) versus nitrogen pressure at $T=\SI{70}{\celsius}$ showing contributions from diffusion to walls (dashed red), Rb-N$_2$ spin-destruction collisions (dashed blue), Rb-Rb spin-exchange (black) and spin destruction (dashed black) collisions. Vertical red line indicates the experimental \SI{3.8}{\torr} of nitrogen pressure and the black star shows the experimental linewidth of $\Gamma\subexp^{\rm{D2}}/2\pi=38$ kHz for the D2 line at $T=\SI{70}{\celsius}$. The red shaded region indicates the pressure range with significant radiation trapping.}
\label{fig:ZFResonances}
\end{figure}

To demonstrate the potential of LWVCs for application to quantum sensors based on atomic coherence, we perform measurements of zero-field magnetic resonance (ZFR) using an elliptically polarized single beam \cite{Shah2009}. The experimental setup is shown in Fig. \ref{fig:ExpSetup} (bottom). The laser beam is partially circularly polarized by a quarter-wave-plate so that the atomic ensemble is optically-pumped \cite{HAPPER1972} with a non-zero electron spin polarization $P_z$ along the z-axis. Then, in the presence of a dc magnetic field applied in the transverse direction $B_x$, the linearly polarized component of the same beam undergoes paramagnetic Faraday self-rotation, which is detected by a polarimeter, consisting of a half-wave-plate, a PBS and a differential photo-detector with switchable gain (Thorlabs PDB450A). For the magnetometry measurements, we use different DFB lasers at either $795$ nm or $780$ nm, tuned near the central D1 or D2 lines of $^{85}$Rb, respectively. In Fig. \ref{fig:ZFResonances}(a-b) we show zero-field magnetic resonances for the D$1$ and D$2$ Rb lines when the transverse magnetic field $B_x$ is scanned over a range of about $100$ $\mu$T. These experimental resonances demonstrate that optical pumping can be performed in the introduced laser-written atomic vapor cells (LWVCs) and that atomic sensors based on atomic spin coherence can be realized with this manufacturing technique, with the potential of integration with laser-written waveguides and PICs.

The physics explaining zero-field magnetic resonances, using a near-resonance single beam with elliptical polarization, is described in \cite{Shah2009} and reported in the Appendix. The detected differential signal is \cite{Shah2009}:
\begin{equation}
V_{\mathrm{diff}}=V_0\sin{\phi}\cos{2\theta},
\label{Eq:signal}
\end{equation}
where $V_0$ is the signal amplitude (in Volt), $\phi\propto P_z$ is the rotation angle,  $\theta$ is the angle of the quarter-waveplate optic axis, relative to the initial linear polarization. In our experiment we fix $\theta=\pi/8$ as optimal tradeoff between pumping and probing \cite{Shah2009}. As described in the Appendix, the equilibrium electron spin polarization:
\begin{equation}
P_z=\frac{P_z^0\Delta B_x^2}{B_x^2+\Delta B_x^2}
\label{eq:EqPol}
\end{equation}
is a Lorentzian function of $B_x$ with half-width-half-maximum (HWHM) given by $\Delta B_x=\Gamma/\gamma$, where the full relaxation rate $\Gamma=R+\Gamma\subdk=1/\tau$ is the inverse of the spin coherence time $\tau$, and it is the sum of the optical pumping rate $R$ and the rate of relaxation in the dark $\Gamma\subdk$, including all spin relaxation effects not caused by the pump/probe beam. This is given by:
\begin{equation}
\Gamma\subdk=\Gamma\subwd+\Gamma\subrt+\Gamma\subcoll,
\label{eq:RelRate}
\end{equation}
where $\Gamma\subwd$ and $\Gamma\subrt$ are the rates due to atomic diffusion to the walls and due to radiation trapping \cite{Rosenberry2007,Seltzer2009}, respectively. The collisional relaxation rate, $\Gamma\subcoll=\Gamma\subbg+\Gamma\subse+\Gamma\subsd$, includes Rb-buffer gas ($\Gamma\subbg$), Rb-Rb spin-exchange  ($\Gamma\subse$) and Rb-Rb spin-destruction ($\Gamma\subsd$) collisional rates, respectively. In our experimental conditions the buffer gas pressure is only $3.8$ Torr so that $\Gamma\subcoll\ll \Gamma\subwd$, $\Gamma\subrt$. Full formulas for all these relaxation rates are given in the Appendix. Since the rotation angle depends linearly on spin polarization,
Eq. (\ref{Eq:signal}) describes a Lorentzian function of $B_x$ when $\phi$ is small, e.g. at low vapor density, and for larger $\phi$ predicts a  ``wrapping around'' of the signal \cite{Shah2009,Seltzer2009}. \\
As shown in Fig. \ref{fig:ZFResonances}(a-b), the amplitude of the self-rotation signal near the D$1$ line, which starts to show a signal "wrap-around" \cite{Shah2009} at 80$^{\circ}$ C, is more than $50\%$ larger than the one near the D$2$ line due to the more efficient optical pumping, i.e. higher achievable spin polarization \cite{SeltzerThesis}, near the D$1$ line. Even at lower temperature, the resonances are not pure Lorentzian functions due to diffusion of polarized atoms in and out of the laser beam, resulting in the Ramsey narrowing effect, previously observed both in EIT \cite{Xiao2006} and magnetometry \cite{Ledbetter2008}. In Fig. \ref{fig:ZFResonances}(c) we show the experimental relaxation rate $\Gamma\subexp/2\pi$ for both D$1$ and D$2$ Rb lines, obtained by fitting the zero-field resonances with Eq. \ref{Eq:signal}. When pumping near the D2 line, as shown in Fig. \ref{fig:ZFResonances}(a), for a number density of $n = 0.74\times 10^{12}$ atoms/cm$^3$, corresponding to a temperature of $T=70^{\circ}$C, the obtained HWHM linewidth is $\Delta B_x^{\rm{exp}}=5.4$ $\mu$T, which gives a full experimental relaxation rate $\Gamma\subexp^\textrm{D2}=\gamma\times\Delta B_x^{\rm{exp}}=2\pi\times38$ kHz. This is in good agreement with the theoretical rate due to collisions with the walls, given by Eq. (\ref{eq:WallsRate}) in the appendix, as shown in Fig. (\ref{fig:ZFResonances}-d) for a N\textsubscript{2} pressure of 3.8 Torr, interaction length $l=9.5$ mm and radius $r=0.6$ mm (where the discrepancy with the $1$ mm $\times$ $1$ mm cross section is due to the cuboid shape of the physics channel). While the contribution due to diffusion to the walls is independent of the degree of atomic polarization, at same temperature we measure a larger broadening $\Gamma\subexp^\mathrm{D1}=2\pi\times82$ kHz for pumping on the D1 line, due to higher pumping rate $R$ and then radiation trapping $\Gamma\subrt$. The latter, given by Eq. (\ref{eq:RadTrap}) in the Appendix, increases rapidly with temperature, due to a nonlinear dependence on Rb number density, and at our buffer gas pressure becomes the dominant broadening mechanism above about \SI{75}{\celsius}. The resulting linewidth is shown in Fig. (\ref{fig:ZFResonances}-c). This is an expected feature due to the ultra low buffer gas pressure associated with the specific filling technique and dispenser activation time. As depicted in Fig. (\ref{fig:ZFResonances}-d) for N$_2$ pressure above 10 Torr, radiation trapping $\Gamma\subrt$ starts to become negligible, as shown in prior works \cite{Rosenberry2007,Seltzer2009}, while the pressure that minimizes the linewidth is obtained as tradeoff between wall collisions $\Gamma\subwd$ and Rb-N\textsubscript{2} spin-destruction collisions $\Gamma\subbg$ \cite{Kitching2018}. In the same figure we also report other contributions from spin-exchange and spin-destruction collisions at $T=70^{\circ}$C. The experimental agreement at low buffer gas pressure allows us to predict the linewidth behaviour at higher pressures giving $\Gamma\subdk^{\rm{th}}=2\pi\times190$ Hz, a two-order-of-magnitude linewidth reduction, for the optimal N$_2$ pressure of about \SI{2472}{\torr}. Together with the absence of radiation trapping, that is a scenario that makes the introduced laser-written vapor cells (LWVCs) attractive for sensitve chip-scale optical magnetometry \cite{Budker2007,Kitching2018} where relaxation due to spin-exchange can be further suppressed in the SERF regime, reaching femtotesla sensitivity \cite{Shah2009}. While high amount of N$_2$ gas can be achieved in miniaturized cells by UV decomposition of rubidium azide (RbN$_3$) \cite{Karlen2017,Tayler2022}, a valuable alternative to significantly decrease both Rb-Rb and Rb-N$_2$ spin destruction collisions is given by the coating of inner walls walls with antirelaxation films such as octadecyltrichlorosilane (OTS) \cite{Seltzer2007} or paraffin \cite{Balabas2010Paraffin,Seltzer2010InvestigationOA}, that have enabled minute-long transverse spin-relaxation time \cite{Balabas2010}. Notably, these anti-relaxation coatings are not compatible with MEMS cells due to high temperature required for anodic bonding, but their application to our LWVCs would be straightforward.

\section{Conclusions}
We have introduced a novel and maskless manufacturing technique of alkali-metal vapor cells based on femtosecond laser writing followed by chemical etching (FLICE). We have demonstrated proof-of-principle applications of a laser-written vapor-cell (LWVC) for sub-Doppler saturated absorption spectroscopy of rubidium and single beam optical magnetometry based on paramagnetic Faraday rotation. The introduced laser-writing technique provides a suitable interaction volume for precision atomic spectroscopy and atomic quantum sensing with 3D structuring versatility and it has high potential for integration with waveguide-based photonic structures \cite{Newman2019,Hummon2018,Sebbag2021} and optical components \cite{Corrielli2014,heil14} as well as with optical fibers for chip-scale \cite{Kitching2018} and fiber-coupled \cite{Simons2018} atomic sensors. Next development steps will focus on the fabrication of sensing chambers with sub-mm cross sections, i.e. in the 100 $\mu$m range, and the interfacing of our LWVC with other components such as GRIN lenses, optical waveguides and fibres, in order to increase the degree of integration and the achievable operational complexity of this tehcnology. As further development, different geometries and filling techniques based on alkali metal azide \cite{Karlen2017}, mixture of argon and nitrogen \cite{mauricePhDThesis} as well as other noble gas species like $^3$He and $^{129}$Xe \cite{Limes2018}, will be implemented to optimize specific applications as atomic references, atomic clocks, optical magnetometers and atomic gyroscopes. 
Finally, the use of NEG dispensers is an attractive feature for miniaturized cold-atoms sensors and compact ultracold quantum technologies \cite{Boudot2020,Little2021,Burrow2021}.  

\section*{Appendix. Theory of zero-field magnetometry resonances}
The physics of optical magnetometry using a single beam with elliptical polarization is described in \cite{Shah2009}. Here we report the functional form used to fit our experimental zero-field magnetic resonances and we give the explicit formulas for the principal relaxation processes.
\subsection*{Equilibrium polarization}
Optical pumping by the circularly polarized component induces non-zero electron spin polarization $\mathbf{P}=\langle\mathbf{S}\rangle/S$ \cite{HAPPER1972} and a change in the refractive index for right/left circular polarization components equal to $n_r=1+k(1+ P_z)\mathrm{Im}[\mathcal{V}(\nu)]/\nu$ and $n_l=1+k(1- P_z)\mathrm{Im}[\mathcal{V}(\nu)]/\nu$, where $\mathcal{V}(\nu)$ is the Voigt profile including natural linewidth, pressure and Doppler broadening, $P_z$ is the spin polarization along the z-axis, $k=nr_ec^2f/4\pi$, where $n$ is the alkali metal number density, $r_e$ is the classical electron radius, and the absorption oscillator strength $f$ is $1/3$ or $2/3$ for the D{1} or D{2} line, respectively. The time-evolution of the spin polarization is described by the Bloch equation \cite{Shah2009}:
\begin{equation}
\dfrac{d\mathbf{P}}{dt}=D\nabla^2\mathbf{P}+\frac{1}{Q(P)}(\gamma\mathbf{P}\times\mathbf{B}+R(\mathbf{s}-\mathbf{P})-\frac{\mathbf{P}}{T_2}),
\label{Eq:Bloch}
\end{equation}
where D is the diffusion constant, $\mathbf{B}$ is the applied magnetic field, $\gamma=2\pi g_s\mu_B/\hbar$ is the electron gyromagnetic ratio, $Q(P)$ is the nuclear slowing down factor \cite{Ledbetter2008}, $\mathbf{s}=-\sin{\theta}\mathbf{\hat{z}}$ is the photon spin, $\theta$ is the angle of the quarter-waveplate optic axis, relative to the initial linear polarization, $R(\nu)=\sigma(\nu)\Phi$ is the optical pumping rate, given by the product of the absortion cross section $\sigma(\nu)=\pi r_ecf\mathrm{Re}[\mathcal{V}(\nu)]$ at given frequency and the photon flux. In the presence of a transverse magnetic field $B_x$, with the other two field components $B_y=B_z=0$, the steady-state solution of Eq. (\ref{Eq:Bloch}) for the z-component of the electron spin polarization is:
\begin{equation}
P_z=\frac{sR/\Gamma}{\gamma^2B_x^2/\Gamma^2+1},
\label{eq:EqPol2}
\end{equation}
which is a Lorentzian function of $B_x$ with half-width-half-maximum (HWHM) given by $\Delta B_x=\Gamma/\gamma$, where $\Gamma=R+\Gamma\subdk=1/\tau$ is the total relaxation rate, $\tau$  is the spin coherence time, $R$ is the optical pumping rate and $\Gamma\subdk$ is the rate of relaxation in the dark. In the main text we report the same Eq. (\ref{eq:EqPol2}) with $P_z^0=sR/\Gamma$.
\subsection*{Diffusion to the walls, radiation trapping and collisional rates}
As described in the main text, in our experimental conditions the dominant relaxation contributions are the rate due to diffusion to the walls and to radiation trapping. The wall relaxation rate for the fundamental diffusion mode of a cylinder is \cite{HAPPER1972,Kitching2018}:
\begin{equation}
\Gamma\subwd=\Big[\Big(\frac{\pi}{l}\Big)^2+\Big(\frac{2.405}{r}\Big)^2\Big]\frac{D_0}{\eta}\sqrt{\frac{T}{273.15}},
\label{eq:WallsRate}
\end{equation}
where $l$ ($r$) is the vapor cell length (radius), $D_0$ is the diffusion constant at \SI{273}{\kelvin} and \SI{760}{\torr} (and thus \SI{1}{\amagat}) and $\eta$ is the nitrogen number density in multiples of one amagat $n_0$. The radiation trapping $\Gamma\subrt$ contributes to spin relaxation due to re-absorption of spontaneously emitted photons and is given by \cite{Seltzer2009}:
\begin{equation}
\Gamma\subrt=K(M-1)QR(1-P_z),
\label{eq:RadTrap}
\end{equation}
where $K$ is the degree of depolarization caused by photon absorption, $M$ is the number of times a photon is emitted before leaving the interaction volume, so that $M-1$ is the average number of times a photon is re-absorbed by atoms after first being scattered from the probe beam. The rate $Q=1/(1+p_Q/p'_Q)$ is the probability for an atom in the excited state to decay by spontaneous emission rather than by quenching, where $p_Q$ is the quenching gas pressure, and $p'_Q$ is the characteristic pressure that gives $Q
=0.5$.
In Eq. (\ref{eq:RadTrap}) $M$ grows with, and consequently $P_z$ decreases with, increasing atomic density, creating a nonlinear dependence of relaxation $\Gamma\subrt$ on number density. \\ To estimate the behaviour at higher buffer gas pressures, where radiation trapping becomes negligible \cite{Rosenberry2007,Seltzer2009}, we calculate the Rb-buffer gas ($\Gamma\subbg$), Rb-Rb spin-exchange  ($\Gamma\subse$) and Rb-Rb spin-destruction ($\Gamma\subsd$) collisional rates. These contributions are given by:
\begin{equation}
\Gamma\subbg=n\sigma_{Rb-N_2}\bar{v}_{Rb-N_2},
\label{eq:buffgas}
\end{equation}
\begin{equation}
\Gamma\subse=q_{SE}n\sigma_{SE}\bar{v}_{Rb-Rb},
\label{eq:se}
\end{equation}
\begin{equation}
\Gamma\subsd=n\sigma_{SD}\bar{v}_{Rb-Rb},
\label{eq:sd}
\end{equation}
where $n=\eta n_0$ is the Rb number density, $\bar{v}_{Rb-N_2}$ ($\bar{v}_{Rb-Rb}$) is the Rb-N$_2$ (Rb-Rb) relative thermal velocity, $q_{SE}=7/32$ is a reduction factor of spin-exchange due to nuclear spin \cite{SeltzerThesis}, $\sigma_{Rb-N_2}=1\times10^{-22}$ cm$^2$, $\sigma_{SE}=1.9\times10^{-14}$ cm$^2$ and $\sigma_{SD}=1.6\times10^{-17}$ cm$^2$ are the Rb-N$_2$, Rb-Rb spin-exchange and Rb-Rb spin-destruction collisional cross-sections \cite{SeltzerThesis}, respectively.
\subsection*{Rotation signal}
Due to the self-induced atomic polarization, the single beam propagating through the atomic medium undergoes paramagnetic Faraday rotation. As described in \cite{Shah2009}, the rotation angle is $\phi=cr_e f n l P_z \mathrm{Im}[\mathcal{V}(\nu)]$, which depends on magnetic field due to Eq. (\ref{eq:EqPol2}), and $l$ is the atomic interaction length. For natural abundance rubidium, the total rotation is the sum of two contributions $\phi_{85}$ and $\phi_{87}$ with relative number density $n_{85}=0.72$ n and $n_{87}=0.28$ n and Voigt profiles for the $^{85}$Rb and $^{87}$Rb isotopes, respectively. When the angle of the half-waveplate in the detection polarimeter of Fig. \ref{fig:ExpSetup} is set to $\alpha=\pi/4$ with respect to the x-polarized light, the detected differential signal, reported in Eq. \ref{Eq:signal} in the main text, is \cite{Shah2009}:
\begin{equation}
V_{\mathrm{diff}}=V_0\sin{\phi}\cos{2\theta},
\label{EqApp:signal}
\end{equation}
where the signal amplitude $V_0$ depends on transmitted power and detector responsivity.

\section*{Funding}
European Regional Development Fund (001-P-001644); H2020 Future and Emerging Technologies (820393, 820405); H2020 Marie Skłodowska-Curie Actions (754510, 766402); H2020 European Research Council (742745); Ministerio de Ciencia, Innovación y Universidades (PGC2018-097056-B-100); Agència de Gestió d'Ajuts Universitaris i de Recerca (2017-SGR-1354); Fundación Cellex; FUNDACIÓ Privada MIR-PUIG; Generalitat de Catalunya (CEX2019-000910-S).

\section*{Acknowledgments}
We thank Charikleia Troullinou and Kostas Mouloudakis for laboratory assistance and useful discussions. We thank Vittoria Finazzi for lending us a UV curing light source and Dr. Luis Guillermo Gerling for the use
of a laboratory glovebox to perform the filling. This project was supported by H2020 Future and Emerging Technologies
Quantum Technologies Flagship projects MACQSIMAL and QRANGE ; H2020 Marie Skłodowska-Curie Actions
projects PROBIST and ITN ZULF-NMR; H2020 European Research Council (ERC) Advanced Grant CAPABLE; Spanish
Ministry of Science project OCARINA and “Severo Ochoa” Center of Excellence CEX2019-000910-S Generalitat de
Catalunya through the CERCA program; Agència de Gestió d’Ajuts Universitaris i de Recerca; Secretaria d’Universitats
i Recerca del Departament d’Empresa i Coneixement de la Generalitat de Catalunya, co-funded by the European Union
Regional Development Fund within the ERDF Operational Program of Catalunya; Fundació Privada Cellex; Fundació
Mir-Puig.

\section*{Disclosure}
The authors declare no conflicts of interest.
\section*{Data availability}
Data underlying the results presented in this paper are not publicly available at this time but may be obtained from the authors upon reasonable request.

\bibliography{LWVC_BA_MI}






\end{document}